\documentclass[11 pt]{article}
\usepackage[a4paper, tmargin=.75in, lmargin=.75in, bmargin=1in, rmargin=.75in]{geometry}
\usepackage[font=scriptsize,labelfont=bf]{caption}
\usepackage{graphicx}   
\usepackage{comment} 
\usepackage{amssymb,amsmath, mathrsfs,mathtools}
\usepackage{setspace}
\usepackage{tabularray}
\usepackage{longtable}
\usepackage{multirow}
\usepackage{array}
\usepackage{bigints}
\usepackage{xurl}
\usepackage{authblk}
\usepackage{lineno}
\usepackage{threeparttable}
\usepackage[sort&compress,numbers]{natbib}
\usepackage[colorlinks,allcolors=blue]{hyperref}

\usepackage{indentfirst}

\providecommand{\keywords}[1]
{\small \textbf{\textit{Keywords---}} #1}

\renewenvironment{abstract}{%
  \textbf\abstractname
  \list{}{\vspace{-0.2 cm} \leftmargin0in \rightmargin\leftmargin}
  \item\relax
}{%
  \endlist \par\bigskip
}

\usepackage{soul} 

\usepackage{titlesec}
\titleformat*{\section}{\large\bfseries}
\titleformat*{\subsection}{\normalsize\bfseries}
\titleformat*{\subsubsection}{\normalsize\it}

\title{\Large{Integrated Population Balance and Multiphysics Modeling for Predicting Undesired Agglomeration in Small Molecule Manufacturing}}

\author[1]{\normalsize Prakitr Srisuma}
\author[1]{\normalsize Peter Hou}
\author[1]{\normalsize Shashank Venkat Muddu}
\author[2]{\normalsize Neda Nazemifard}
\author[1]{\normalsize Allan S. Myerson}
\author[1\footnote{Corresponding author. Email: braatz@mit.edu}]{\normalsize Richard D. Braatz}

\affil[1]{\scriptsize Massachusetts Institute of Technology, Cambridge, MA 02139, USA} 
\affil[2]{\scriptsize Takeda Development Center Americas, Inc., Cambridge, MA 02139, USA}

\date{}

\begin{document}
\maketitle

\begingroup
\parindent=0cm
\parskip=0.25cm

\hrule

\noindent\begin{abstract}
Agitated filter dryers (AFDs) are a crucial unit operation in small molecule manufacturing that enables simultaneous filtration, washing, and drying of active pharmaceutical ingredients. One of the key challenges in AFDs is associated with undesired agglomeration, where the presence of hard agglomerates results in off-spec products, equipment damage, and additional downstream processing. This article presents a novel mechanistic model that describes the formation of soft and hard agglomerates during agitated filter drying. By integrating population balance and multiphysics modeling, the model can accurately predict the evolution of the product temperature, moisture content, and particle size distribution, and hence quantify the extend and impact of undesired agglomeration across various operating conditions. Our proposed model-based framework enables the rational design and operation of AFDs for improving the product quality and process reliability. 
\end{abstract}
\vspace{-0.5 cm}

\vspace{0.4 cm}
\hrule
\vspace{0.4 cm}
\endgroup

\normalsize

\section{Introduction} \label{sec:Intro}
Agitated filter dryers (AFDs) are an important unit operation in small molecule manufacturing, integrating filtration, washing, and drying within a single vessel. The main objective of AFDs is to recover purified, dry solid crystals (active pharmaceutical ingredients, APIs) from a slurry after crystallization. In AFDs, solid particles are continuously heated and agitated under partially wet conditions. This agitation can help enhance heat and mass transfer in the process, but can also promote particle agglomeration, potentially causing significant changes in the particle size distribution (PSD) of the product \cite{Gnanenthiran2025AFDReview,Sahni2013AFDPerformance}. 

Agglomerates are typically classified into two categories: (1) soft agglomerates, which can spontaneously deagglomerate into primary particles during processing, and (2) hard agglomerates, which require significant external forces to break \cite{Nichols2002AgglomerateAggregate}. Hard agglomerates are undesired because they irreversibly change the particle size distribution (PSD) of the product, which results in off-spec products, equipment damage, and additional downstream processing \cite{Gnanenthiran2025AFDReview,Tamrakar2016Dynamic}. Undesired agglomeration is considered the key challenge in AFDs, and thus it is crucial to avoid or minimize the formation of hard agglomerates during the process.

Several experimental studies have investigated how operating conditions and material properties influence particle agglomeration during agitated filter drying. Some important parameters include the API solubility, agitation speed and duration, drying temperature and pressure, initial solvent content, and particle morphology  \cite{Kougoulos2011AgitatedDrying,Sahni2013AFDPerformance,Tamrakar2016Dynamic,Lim2016FilterAgglomeration,Ottoboni2020Experiment}. However, While AFDs have been studied and used widely, the mechanisms underlying undesired agglomeration are not clearly understood \cite{Gnanenthiran2025AFDReview}. Consequently, the design and operation of AFDs still mostly rely on experimental data. We refer to \cite{Gnanenthiran2025AFDReview} for a detailed review of previous studies on AFDs.

Computational studies supporting model-based design and control of AFDs are relatively limited. In particular, no existing mathematical model can predict the formation of hard and soft agglomerates. Some available models focus solely on heat and mass transfer modeling while neglecting particle agglomeration \cite{Nere2012DryingOptimization,Belekar2022AFDModel}, which could be useful for process design and optimization when agglomeration is not significant. A number of existing models rely on the discrete element method (DEM) to understand and predict the particle-level dynamics, e.g., velocity profiles, particle motion, and interactions between particles \cite{Sahni2012DryingPerformance,Sahni2013ContactDryingModel,Tamrakar2020DEM,Irndorfer2026DigitalTwin}. Nevertheless, such models were not developed to predict the evolution of PSD or describe the formation of hard and soft agglomerates. Population balance modeling (PBM) provides a more direct framework for describing particle agglomeration and size evolution, but its direct application to AFDs has not been studied extensively. One recent example is the study by \cite{Togni2025Attrition}, which developed a two-dimensional population balance model describing the attrition of particles for a system of L-Threonine and ethanol during agitated filter drying. More commonly, PBM has been used to model particle agglomeration and breakage in some similar systems such as granulation \cite{Ouchiyama1974Granulation,Liu2018PredictivePBM,Hayashi2019PBM}. A more complicated strategy coupling PBM and DEM has also been investigated \cite{Barrasso2015PBMDEM,Hayashi2020PBMDEMCFD}. Despite a wide range of modeling strategies, none of the aforementioned models can predict the particle size distribution while explicitly distinguishing between the formation of hard and soft agglomerates. Therefore, they cannot fully describe the development of undesired agglomeration, which is the main concern in AFDs.

This article presents a novel mechanistic model that describes the formation of soft and hard agglomerates in AFDs. A combination of population balance and multiphysics modeling is used to capture the complex coupled phenomena that involve heat transfer, mass transfer, and particle interactions, including agglomeration and breakage. The model is evaluated using two experimental systems representing distinct agglomeration behaviors: a limiting case in which hard agglomerates are absent and a practical case in which hard agglomerates form. The validated model is then employed for the analysis, design, and optimization AFDs.

This article is organized as follows. Section \ref{sec:Experiment} describes the experimental setup and details the process of interest. Section \ref{sec:Modeling} presents the development of our mechanistic model for agitated filter drying. Section \ref{sec:Results} showcases various model validation results and simulation studies. Finally, Section \ref{sec:Conclusion} summarizes the study.

\section{Experimental Setup and Process Description} \label{sec:Experiment}

\begin{figure}[ht!]
\centering
    \includegraphics[scale=.52]{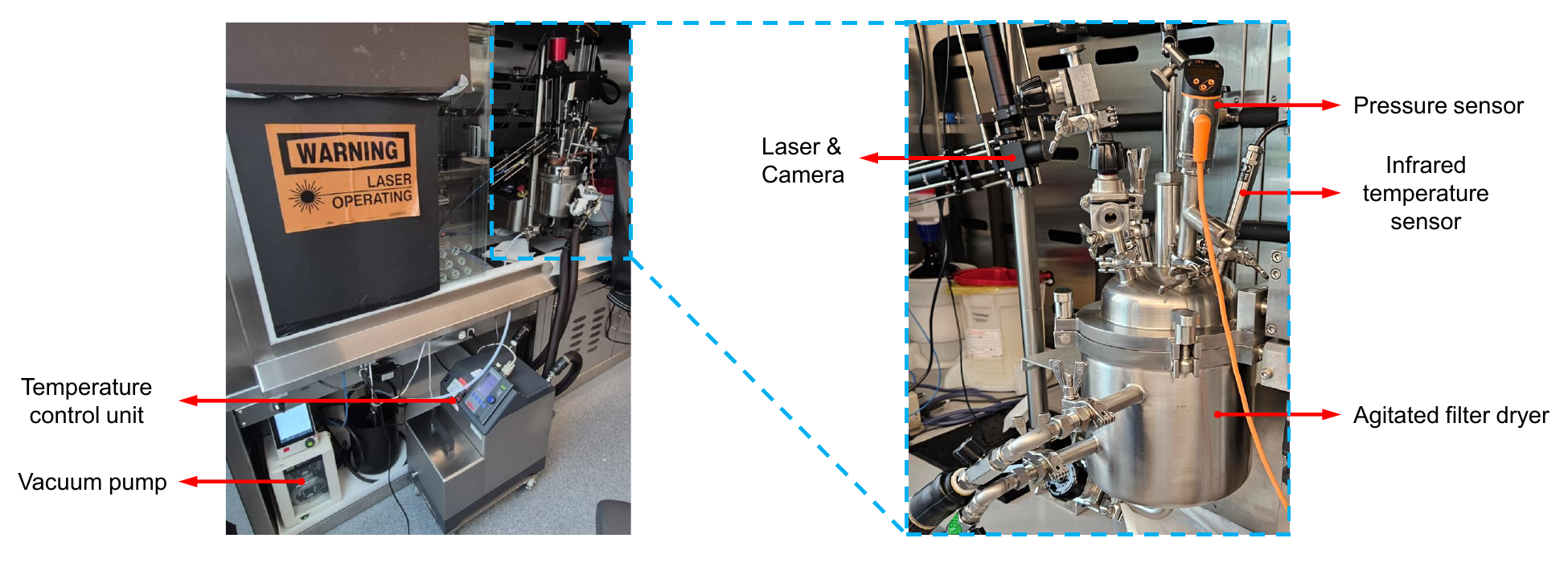}
    \caption{Our agitated filter dryer (AFD) experimental setup showing the overall system, AFD vessel, and sensors.}
    \label{fig:AFD_System}
\end{figure}

The AFD system used in our experiment consists of a dryer (GFD Lab, PSL), a jacketed filter basket (10 $\mu$m mesh), a lid equipped with an agitator motor, an infrared temperature sensor (PyroEpsilon), a pressure sensor (PN2694, ifm), and a sight glass for laser-speckle particle size distribution (PSD) monitoring (Fig.~\ref{fig:AFD_System}). The system was integrated with a temperature control unit (Huber Ministat 125) and a vacuum pump (Büchi V-300). All components were operated and monitored through a custom Python-based human–machine interface (HMI) running on a computer. For standard operations, During standard operation, the powder was loaded into the basket and soaked in a solvent introduced through the lid inlet port while being agitated at 5 rpm for 15 min. The solvent was then drained through a bottom valve into a separator. Following this initial filtration step, the jacket temperature, vacuum pressure, and agitator speed were configured on the HMI to initialize the automated drying and data-logging sequence.

The operating temperature and pressure were continuously monitored using inline instrumentation. The sensors, mounted directly on the AFD lid, can measure the internal process conditions in real time throughout the drying cycle. The infrared sensor provided non-contact measurements of the powder-bed surface temperature by detecting the naturally emitted thermal radiation and focusing it onto an internal infrared detector. The resulting temperature signal was converted into a robust 4–20 mA analog current output and transmitted to the computer. The custom HMI continuously recorded the temperature profile and vacuum pressure, enabling verification of stable operating conditions throughout the drying process.

Real-time physical attribute changes and crystalline transitions were tracked non-invasively using an inline optical setup. The dryer lid was equipped with a specialized sight glass allowing a laser beam to illuminate the powder bed for laser-speckle-based PSD monitoring. The backscattered light from the moving powder surface was collected continuously during the drying process, and the raw speckle images were converted into transient PSD profiles via the previously published Physics-Enhanced Autocorrelation-based Estimator (PEACE) framework \cite{zhang2023extracting}. The drying process was automatically terminated once the estimated PSD trend stabilized, with stabilization defined as a period during which the mean particle size did not fluctuate by more than 5\% for at least 5 min.


To monitor the moisture (solvent) content, gravimetric sampling was used. At the determined endpoint of a standard run, a 15–20 g sample was collected, weighed, and dried for 24 h in a vacuum oven at a solvent-dependent temperature to determine the final residual solvent content. Additionally, to obtain the time profile of the moisture content, the system was programmed to automatically pause every 30 min during the drying cycle to allow the extraction of a smaller 5–10~g sample. Each sample was immediately weighed, dried overnight under identical vacuum oven conditions to ensure complete volatile removal, and reweighed to calculate its moisture content.

\section{Mechanistic Modeling} \label{sec:Modeling}
\begin{figure}[ht!]
\centering
    \includegraphics[scale=.37]{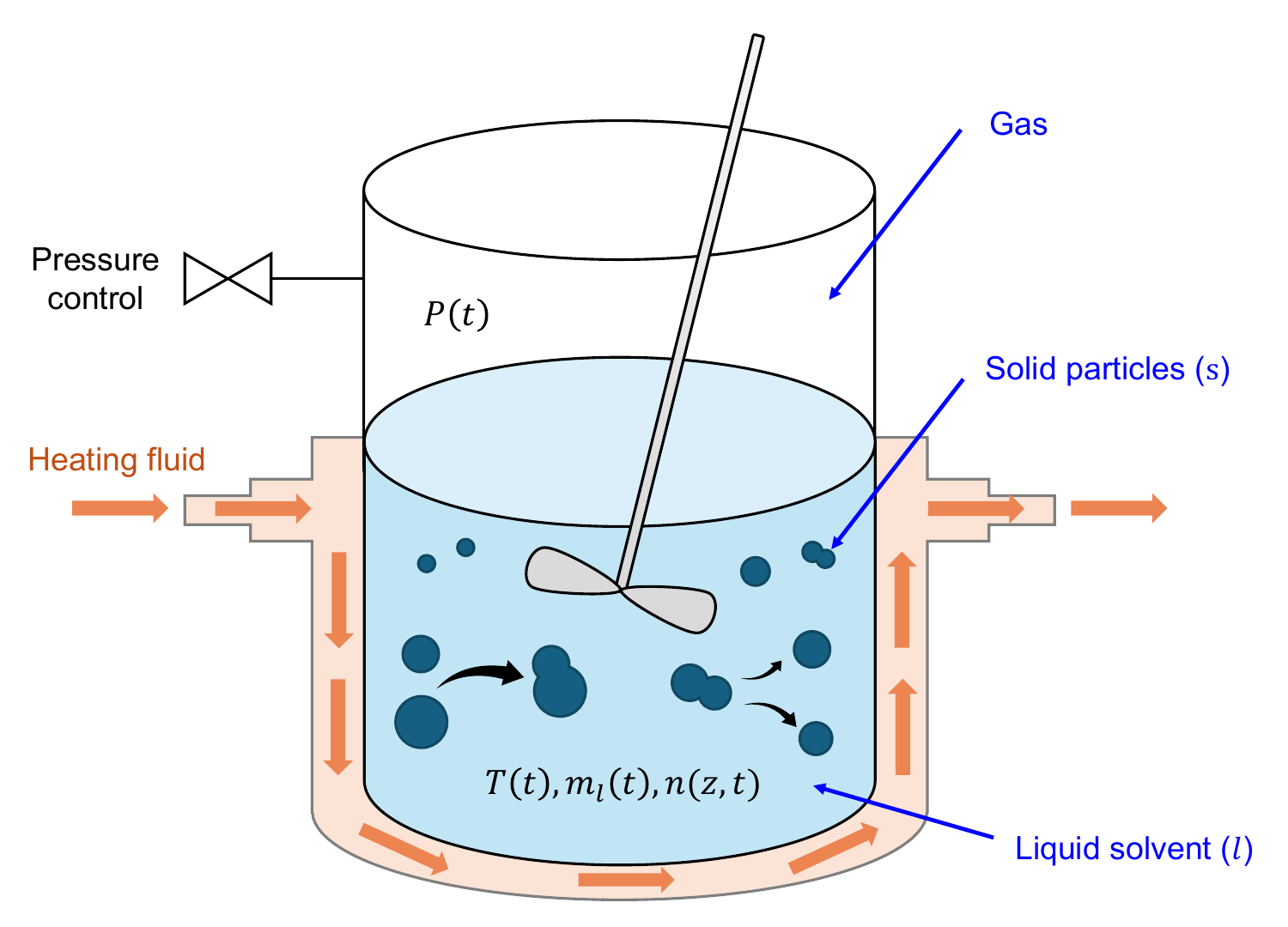}
    \caption {Schematic diagram showing the mechanistic modeling of the agitated filter drying process, where the system temperature, solvent content, and number of particles are denoted by $T$, $m_l$, and $n$, respectively.} 
    \label{fig:Schematic}      
\end{figure}

Our experimental setup (Fig.~\ref{fig:AFD_System}) is represented by a schematic diagram in Fig.~\ref{fig:Schematic}. The system consists of the solid powders (e.g., Aspirin) and liquid solvent (e.g., water). The system is heated via the heating fluid (e.g., hot oil) fed through the well-insulated jacket. The operating pressure can be manipulated by adjusting the amount of inert gas but is typically kept constant most of the time. Following Fig.~\ref{fig:Schematic}, our mechanistic model focuses on describing three main aspects of AFDs, namely (1) heat transfer, (2) mass transfer, and (3) particle agglomeration and breakage.

\subsection{Heat transfer} \label{sec:Model-HeatTransfer}
The first part of the model describes heat transfer during agitated filter drying. Since the solid particles and liquid solvent are agitated continuously, it is assumed that the system is well mixed. Heat transfer in the system is mainly associated with the heating fluid, inert gas, irreversible degradation of mechanical (due to agitation) to internal energy, and evaporative cooling. Evaporative cooling is a simultaneous heat and mass transfer process, which is discussed further in Section \ref{sec:Model-MassTransfer}.  

For heat transfer, the energy balance equation is
\begin{equation} \label{eq:energy}    (m_sC_{p,s}+m_lC_{p,l})\frac{dT}{dt} = U_h A_h(T_h-T) +U_g A_g(T_g-T) + W_\omega + \Delta H_\textrm{vap}\frac{dm_l}{dt},  
\end{equation}
where $T$ is the product temperature, $T_h$ is the heating fluid temperature, $T_g$ is the gas temperature, $t$ is time, $m$ is the total mass, $C_p$ is the heat capacity, $U$ is the overall heat transfer coefficient, $A$ is the heat transfer area, $W_\omega$ is the shaft work associated with agitation, $\Delta H_\textrm{vap}$ is the heat of vaporization, and the subscripts `$s$', `$l$', `$h$', `$g$' denote the solid particles, liquid solvent, heating fluid, and gas, respectively. In this case, the temperature, solvent mass, and heat transfer area are time dependent. Since the total mass of solid particles does not change with time, $m_s$ does not change and remains at its initial value at all times. Other parameters are assumed to be constant. 

The initial condition for \eqref{eq:energy} is
\begin{equation}
    T(t_0) = T_0
\end{equation}
where $t_0$ is the initial time.

\subsection{Mass transfer} \label{sec:Model-MassTransfer}
Mass transfer in AFDs primarily entails the drying process, i.e., solvent evaporation. The driving force for evaporation depends on the characteristics of drying, namely (1) constant-rate drying and (2) falling-rate drying \cite{Seader2007Separation}. In the constant-rate drying regime, the solid surfaces remain saturated with the liquid solvent, where the evaporation rate is controlled primarily by external mass and heat transfer between the liquid and environment (gas). In the falling-rate drying regime, some of the solid surfaces are not entirely covered by the liquid, and so the rate of drying is limited by internal transport of the solvent within the solid. Both drying regimes require different modeling strategies.

For the constant-rate drying regime, the liquid evaporation rate at the surface is \cite{Mills1995HeatTransfer,Seader2007Separation}
\begin{equation} \label{eq:evap_constantrate}
    \frac{dm_l}{dt} = -h_m A_g(x_{\textrm{sat}}-x_{e}),
\end{equation}
where $h_m$ is the mass transfer coefficient, $x_{\textrm{sat}}$ is the mass fraction of the liquid at the liquid-vapor interface (equilibrium), and $x_{e}$ is the mass fraction of the liquid in the gas phase (environment). The subscripts `sat' and `$e$' here denote the equilibrium condition and environment, respectively. By assuming the ideal gas law, the mass fraction can be calculated by \cite{Mills1995HeatTransfer}
\begin{gather}
    x_{\textrm{sat}} = \frac{p_{\textrm{sat}}M_l}{p_{\textrm{sat}}M_l + (P - p_{\textrm{sat}})M_g}, \\
    x_{e} = \frac{p_{e}M_l}{p_{e}M_l + (P - p_{e})M_g},   
\end{gather}
where $P$ is the total pressure, $p$ is the partial pressure of the solvent, and $M$ is the molar mass. The saturation pressure $p_{\textrm{sat}}$ is a function of temperature. For water \cite{Smith2018Thermo},
\begin{equation} \label{eq:psat}
    p_{\textrm{sat}} = 10^3\exp\!\left(\!16.3872-\frac{3885.7}{T-42.98}\right).
\end{equation}
To summarize, in the constant-rate drying regime, the driving force for mass transfer is driven by the difference between the amount of the liquid solvent at the equilibrium condition and that in the gas phase (environment).

For the falling-rate drying regime, the governing transport process entails internal diffusion of the liquid through the solid structures. Instead of using a full partial differential equation (PDE) for mass transport, a semi-theoretical model that is much simpler but also provides accurate prediction is \cite{Eldeen1979FallingRateDrying,Ertekin2017ThinLayerReview,Siles2018DryingModel}
\begin{equation} \label{eq:evap_dryingrate}
    \frac{dX}{dt} = -k_m(X-X_{\infty}),
\end{equation}
where $X=m_l/m_s$ is the mass fraction of the liquid solvent in the solid particles, $X_{\infty}$ is the equilibrium concentration, and $k_m$ is the rate constant for diffusion. Multiplying \eqref{eq:evap_dryingrate} by $m_s$ results in the mass transfer rate for \eqref{eq:energy}. In this case, the mass transfer process and related driving force are associated with the solid-vapor interface. Notably, this is different from the constant-rate drying regime, where the transport process occurs at the liquid-vapor interface. 

The initial conditions \eqref{eq:evap_constantrate} and \eqref{eq:evap_dryingrate} are
\begin{gather}
    m_l(t_0) = m_{l0}, \\
    X_l(t_0) = X_{l0}.
\end{gather}

\subsection{Particle agglomeration and breakage} \label{sec:Model-PBM}
During agitated filter drying, some particles may agglomerate to form larger particles, while those large particles can also break (aka deagglomerate), resulting in smaller particles. As a result, the PSD can evolve over time. This final component of the model focuses on describing these phenomena using population balance modeling. 

\subsubsection{Population balance modeling} \label{sec:Model-PBM-basic}
The population balance equation consdering agglomeration and breakage is
\begin{equation}
\frac{\partial n(z,t)}{\partial t} = S_a^{+}(z,t) - S_a^{-}(z,t) + S_b^{+}(z,t) - S_b^{-}(z,t),
\end{equation}
where $n$ is the number density of the solid particles, $z$ is the particle diameter (i.e., internal coordinate of the population balance equation), $S_a$ is the agglomeration term, $S_b$ is the breakage term, and the symbols `+' and `-' denote the birth and death terms, respectively. The agglomeration and breakage terms are \cite{Pena2017SphericalAgglomeration,Bertin2016Numerical}
\begin{gather}
S_a^{+}(z,t) = \frac{z^{2}}{2} \int_{0}^{z} \frac{ \beta\!\left( (z^{3}-\lambda^{3})^{1/3},\, \lambda \right) \, n(z^{3}-\lambda^{3},t)\, n(\lambda,t)
}{(z^{3}-\lambda^{3})^{2/3}}\, \mathrm{d}\lambda, \\
S_{a}^{-}(z,t) = n(z,t)\int_{0}^{\infty}\beta(z,\lambda)\,n(\lambda,t)\,\mathrm{d}\lambda, \\
S_{b}^{+}(z,t) = \int_{0}^{\infty} \nu(\lambda)\, b(\lambda)\, \mathbb{P}(z,\lambda)\, n(\lambda,t)\,\mathrm{d}\lambda, \\
S_{b}^{-}(z,t) = b(z)\, n(z,t),
\end{gather}
where $\beta$ is the agglomeration kernel, $b$ is the breakage kernel, $\nu$ is the average number of particles resulting from breakage, and $\mathbb{P}$ is the breakage probability function. Note that the above derivation assumes that the kernels and probability function are size-dependent. In the later sections, these parameters are defined more precisely for our model and the AFD system.

\subsubsection{Formation of soft and hard agglomerates} \label{sec:Model-PBM-Agglomerate}
The set of equations described in Section \ref{sec:Model-PBM-basic} is sufficient to describe a system that consists of only one particle type. However, in this case, the AFD system comprises (1) soft agglomerates and (2) hard agglomerates. Therefore, additional equations are needed. Given that the subscripts $\sigma$ and $\eta$ denote the soft and hard agglomerates, respectively, the population balance equations for both particle types are
\begin{gather}
\frac{\partial n_\sigma(z,t)}{\partial t} = S_{a,\sigma}^{+}(z,t) - S_{a,\sigma}^{-}(z,t) + S_{b,\sigma}^{+}(z,t) - S_{b,\sigma}^{-}(z,t) - r_{\sigma\rightarrow\eta}(z,t), \label{eq:soft_PBM} \\
\frac{\partial n_\eta(z,t)}{\partial t} = r_{\sigma\rightarrow\eta}(z,t), \label{eq:hard_PBM}
\end{gather}
where $r_{\sigma\rightarrow\eta}$ is the rate of conversion from soft to hard agglomerates, which is assumed to follow a first-order process
\begin{equation}
    r_{\sigma\rightarrow\eta}(z,t) = k_\textrm{c}n(z,t),
\end{equation}
where $k_\textrm{c}$ is the rate constant (defined later in this section). In our model, we assume that hard agglomerates cannot agglomerate or deagglomerate further. The initial conditions are
\begin{gather}
    n_\sigma(z,t_0) = n_{\sigma0}(z), \\
    n_\eta(z,t_0) = 0,
\end{gather}
indicating that there is no hard agglomerate at the beginning of the process.

\begin{figure}[ht!]
\centering
    \includegraphics[scale=.5]{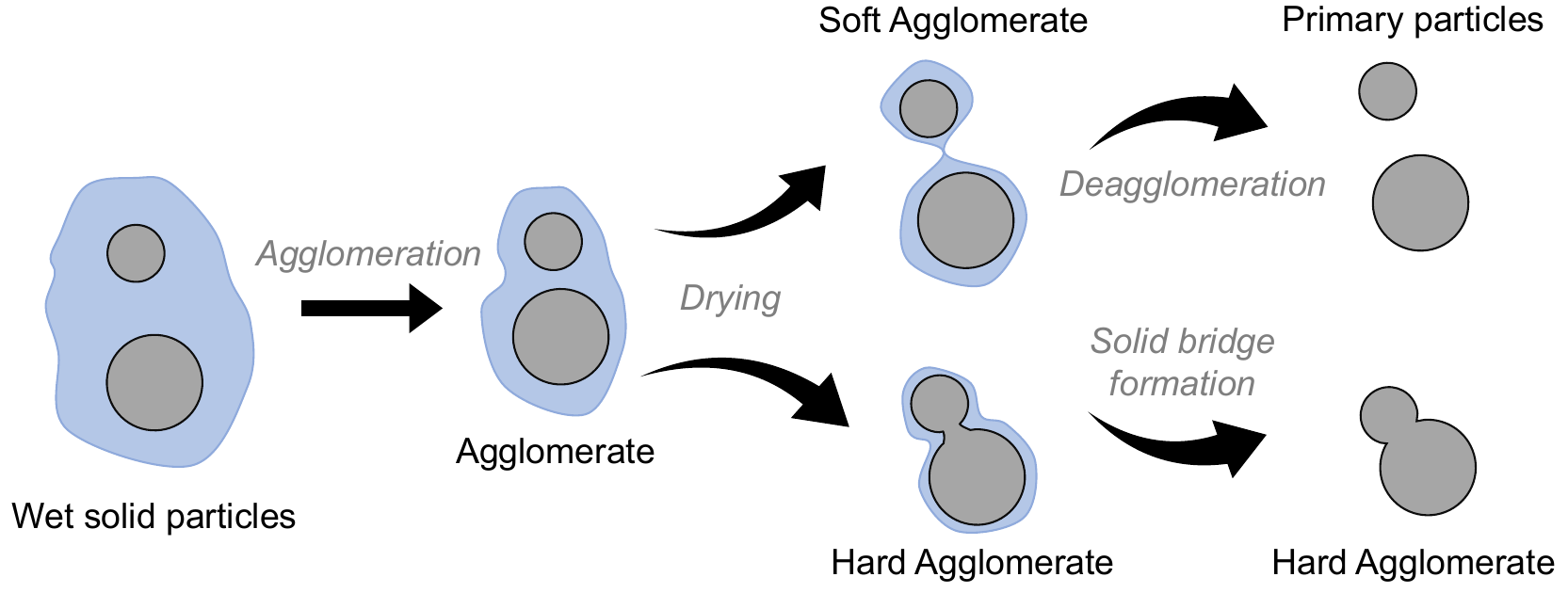}
    \caption {Schematic diagram showing the formation of hard and soft agglomerates during the agitated filter drying process.} 
    \label{fig:Agglomeration}      
\end{figure}

The key phenomena underpinning the formation of soft and hard agglomerates in AFDs are summarized in Fig.~\ref{fig:Agglomeration}. At the beginning of the process, a mixture of liquid (solvent) and solid particles (APIs) from the upstream process, e.g., crystallization, are fed to the AFD chamber. The liquid solvent acts a binding medium that loosely attracts several particles together, facilitating particle agglomeration. Therefore, agglomeration predominantly occurs at the early stage of the drying process when the amount of liquid (i.e., moisture content, solvent content) is high. As the drying process progresses, the liquid solvent evaporates, and thus the moisture content decreases. When the system transitions from the wet regime to the dry regime, there are two competitive phenomena that define the formation of soft and hard agglomerates. First, the liquid bridges that loosely bind multiple solid particles together disappear due to evaporation. Second, the solute that dissolves in the liquid solvent crystallizes out of the liquid phase. As a result of crystallization, the crystals can form a solid bridge attracting the solid particles together more firmly. If evaporation is sufficiently fast that the liquid bridge disappears before the solid bridge can form successfully, then the particles deagglomerate into their original forms (primary particles), in which we classify this type of agglomerates as soft agglomerates. On the other hand, if the growth rate (crystallization) is sufficiently high that the solid bridge can form successfully and bind the particles together firmly, then those particles become hard agglomerates. Soft agglomerates naturally break and deagglomerate back into their original forms during the agitated filter drying process, hard agglomerates are much more difficult to break, requiring considerable external forces. Consequently, the formation of hard agglomerates is undesired as it can irreversibly change the PSD during the process, potentially leading to off-spec products, equipment damage, and additional downstream processing.

Given the aforementioned mechanistic understanding, we define $k_\textrm{c}$ as
\begin{gather}
k_\textrm{c} = f_1\left(\frac{\dot{V}_s}{\dot{V}_l}\right),
\end{gather}
where $f_1$ is a monotonically increasing function (see Table \ref{tab:functions}, $\dot{V}_l$ is the liquid evaporation rate (liquid bridge disappearance), and $\dot{V}_s$ is the growth rate (solid bridge formation). The evaporation rate can be calculated from \eqref{eq:evap_constantrate} or \eqref{eq:evap_dryingrate}, that is,
\begin{equation}
    \dot{V}_l = -\frac{1}{\rho_l}\frac{dm_l}{dt}.
\end{equation}
The growth rate follows a typical relation
\begin{equation}
    \dot{V}_s = k_G\left(C-C^*\right)^b,
\end{equation}
where $k_G$ is the growth constant, $b$ is the growth order, $C$ is the solute concentration (that is dissolved in the solvent), and $C^*$ is the solubility. The crystal growth rate is conventionally expressed as length per unit time. In this case, however, it is converted to a volumetric basis (volume per unit time) to directly compare with the evaporation rate, and thus the geometric parameters are incorporated in $k_g$. The concentration can be calculated by
\begin{equation}
    C = \frac{m_d}{m_l},
\end{equation}
where $m_d$ is the total mass of the dissolved solute, which must satisfy the mass balance
\begin{equation}
    \frac{dm_d}{dt} =  -\dot{V}_s,
\end{equation}
with the initial condition
\begin{equation}
    m_d(t_0) =  m_{d0}.
\end{equation}

\subsubsection{Agglomeration and breakage kernels} \label{sec:Model-PBM-Kernels}
There are various types of agglomeration and breakage kernels in the literature, including those derived from first principles and empirical correlations. The kernels can be dependent on various factors, such as operating conditions (e.g., temperature), fluid properties (e.g., viscosity), and particle sizes; we refer to \cite{Patruno2009BreakageKernels,Pena2017SphericalAgglomeration} for more detailed discussion. Kernels with more parameters can generally capture a wider range of physical behaviors, potentially improving model generalizability across different systems. However, this generalizability often comes at the cost of identifiability and overparameterization. Our goal is to choose the kernels that can sufficiently describe the AFD system and phenomena of interest while keeping the number of parameters minimum. Given the phenomena and mechanistic understanding described in Section \ref{sec:Model-PBM-Agglomerate}, we define the agglomeration and breakage kernels to be a function of the moisture content. Since agglomeration tends to occur better at a higher moisture content, 
\begin{equation}
    \beta = f_2\left(\frac{m_l}{m_s}\right),
\end{equation}
where $f_2$ is a monotonically increasing function. On the other hand, deagglomeration occurs better when the moisture content is low, that is,
\begin{equation}
    b = f_3\left(\frac{m_l}{m_s}\right),
\end{equation}
where $f_3$ is a monotonically decreasing function. Both $f_2$ and $f_3$ are defined in Table \ref{tab:functions}.

\begin{table}[ht!]
\renewcommand{\arraystretch}{1.2}
\caption{Semi-empirical correlations used in the model.} 
\label{tab:functions}
\centering
\begin{threeparttable}
\renewcommand{\arraystretch}{1.2}
\begin{tabular}{@{}m{0.1\textwidth} m{0.32\textwidth} m{0.52\textwidth}@{}}
\hline
\textbf{Notation} & \textbf{Expression} & \textbf{Description} \\
\hline
$f_1(x)$ &  $Kx$ & Conversion from soft agglomerates to hard agglomerates is assumed to follow a simple first order process, where $K$ is a constant. \\

$f_2(x)$& 
$\beta_0\left(\left(\dfrac{x}{x_\textrm{lim}}\right)^{c_1}-1\right), \quad x \geq x_\textrm{lim}$ 
& A simple power law is assumed for the agglomeration kernel, where $c_1$ and $\beta_0$ are constants. When the moisture level is too low, i.e., below $x_\textrm{lim}$, agglomeration is assumed to be negligible, that is, $f_2(x)=0$ for $x<x_\textrm{lim}$.   \\

$f_3(x)$& 
$b_0\left(1-\left(\dfrac{x}{x_\textrm{lim}}\right)^{c_2}\right), \quad x \leq x_\textrm{lim}$ 
& A simple power law is assumed for the breakage kernel, where $c_2$ and $b_0$ are constants. When the moisture level is too high, i.e., above $x_\textrm{lim}$, breakage is assumed to be negligible, that is, $f_3(x)=0$ for $x > x_\textrm{lim}$. \\

\hline  
\end{tabular}
\renewcommand{\arraystretch}{1}
\end{threeparttable}
\end{table}

\subsubsection{Breakage probability function} \label{sec:Model-PBM-Prob}
In general, the breakage term has one additional parameter besides the kernel, namely the breakage probability function $\mathbb{P}(z,\lambda)$. This parameter defines the distribution of smaller particles resulting from deagglomeration. Specifically, it describes the number density of particles of size $z$ resulting from the breakage of particles of size $\lambda$. Considering the fact that soft agglomerates tend to deagglomerate back into their primary particles (i.e., initial conditions) as explained in Section \ref{sec:Model-PBM-Agglomerate}, the breakage probability distribution is assumed to follow the initial conditions $n_{\sigma0}(z)$.

\subsection{Gas dynamics} \label{sec:Model-Gas}
This part of the model is optional and should be considered only when the operating pressure changes significantly over time.The operating pressure in AFDs can be varied by manipulating the amount of inert gas in the system, which can influence heat and mass transfer in the process.

By assuming adiabatic expansion and compression of an ideal gas, the evolution of the inert gas temperature is
\begin{equation}
    \frac{dT_g}{dt} = \frac{\gamma-1}{\gamma}\frac{T_g}{P}\left(\frac{dP}{dt}+\frac{U_g A_g(T-T_g)}{V_g}\right),
\end{equation}
where $T_g$ is the gas temperature, $P$ is the total pressure, and $\gamma$ is the ratio of heat capacities. In general, the pressure profiles $P$ and $dP/dt$ are specified, and the gas temperature $T_g$ can be calculated consequently.

\subsection{Model summary and numerical methods} \label{sec:Model-Summary}
\begin{figure}[ht!]
\centering
    \includegraphics[scale=.45]{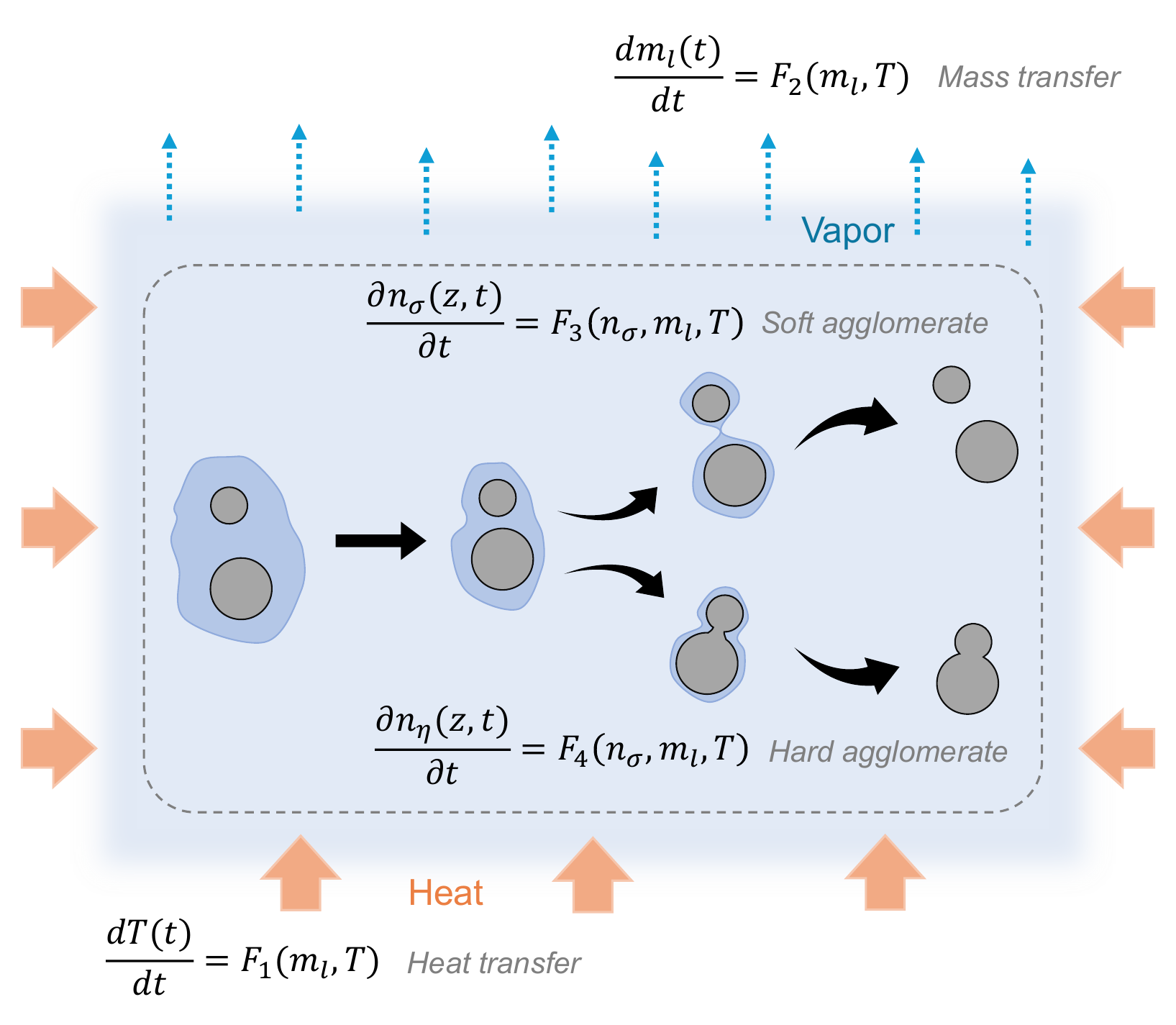}
    \caption {Schematic diagram describing the proposed mechanistic model. The nonlinear functions $f_1$, $f_2$, $f_3$, and $f_4$ are derived from energy, mass, and population balances.} 
    \label{fig:Models}      
\end{figure}

With the modeling strategies described in Sections \ref{sec:Model-HeatTransfer}, \ref{sec:Model-MassTransfer}, and \ref{sec:Model-PBM}, our final model focuses on four important states, namely the product temperature ($T$), moisture content ($m_l$), number density of soft agglomerates ($n_\sigma$), and number density of hard agglomerates ($n_\eta$). The operating pressure ($P$) and inert gas temperature ($T_g$) could be of interest when the pressure varies significantly throughout the process. Our model is pictorially summarized in Fig.~\ref{fig:Models}.

The key model equations include \eqref{eq:energy} for temperature calculation, \eqref{eq:evap_constantrate} or \eqref{eq:evap_dryingrate} for moisture content calculation, and \eqref{eq:soft_PBM} and \eqref{eq:hard_PBM} for number density calculation. The ordinary differential equations (ODEs) \eqref{eq:energy}, \eqref{eq:evap_constantrate}, and \eqref{eq:evap_dryingrate} can be integrated directly using a proper ODE solver. The partial differential equations (PDEs) \eqref{eq:soft_PBM} and \eqref{eq:hard_PBM} are discretized using the finite volume method, resulting in a system of ODEs 
\begin{gather}
\frac{d n_{\sigma,i}}{dt} = S_{a,\sigma,i}^{+} - S_{a,\sigma,i}^{-} + S_{b,\sigma,i}^{+} - S_{b,\sigma,i}^{-} - r_{\sigma\rightarrow\eta,i}, \label{eq:soft_PBM_discretized} \\
\frac{d n_{\eta,i}}{dt} = r_{\sigma\rightarrow\eta,i}, \label{eq:hard_PBM_discretized}
\end{gather}
where $i$ is the integer index that represents a bin for particles with size $z_i$. The resulting system of ODEs can then be integrated using an ODE solver. 

In this work, all simulations were performed in MATLAB R2025b on a computer equipped with an AMD\textsuperscript{\textregistered} Core\texttrademark\ 
Ryzen AI 9 HX 370 processor (12 cores) and 32 GB RAM running 64-bit Windows 11. All parameter and experimental date can be found in the provided software package (see the Data and Code Availability section).

\section{Results and Discussion} \label{sec:Results}
\subsection{Model validation} \label{sec:ModelValidation}
The mechanistic model proposed in Section \ref{sec:Modeling} is validated with a set of experimental data obtained from the equipment described in Section \ref{sec:Experiment}. In this study, our model validation considers two distinct AFD scenarios. The first case, denoted as Case 1, considers a mixture of potassium chloride (KCl) and hexane. Case 1 represents a limiting case in which only soft agglomerates exist because the solid (KCl) is completely insoluble in the solvent (hexane). Hence, no hard agglomerate can form due to the lack of solid bridge formation. The second case, denoted as Case 2, considers a mixture of aspirin (ASA) and isopropyl alcohol (IPA). Case 2 represents a more practical case in which both soft and hard agglomerates can form as ASA is highly soluble in IPA.

\begin{figure}[ht!]
\centering
    \includegraphics[scale=1]{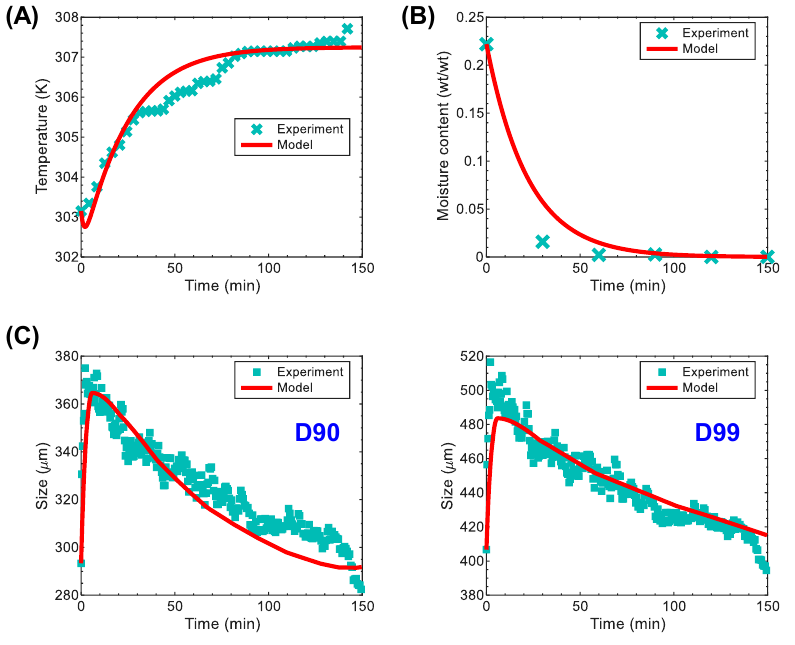}
    \caption {Comparison between the experimental data and model prediction for the system consisting of KCl and hexane (Case 1). Panel (A) shows the product temperature. Panel (B) shows the moisture content. Panel (C) shows D90 and D99.} 
    \label{fig:Exp59_Validation}      
\end{figure}

For Case 1 (KCl + hexane), the model predictions agree very well with the experimental data for all variables of interest, including the temperature, moisture content, and particle size (Fig.~\ref{fig:Exp59_Validation}). The product temperature increases throughout the process and eventually approaches the hot-oil temperature, following the driving force for heat transfer (Fig.~\ref{fig:Exp59_Validation}A). The residual moisture decreases relatively rapidly during the early stage of the process because the evaporation rate is high (Fig.~\ref{fig:Exp59_Validation}B). As drying progresses, the drying rate gradually decreases as the driving force for mass transfer diminishes, until the residual moisture approaches the target value.

Regarding particle size, the formation of hard agglomerates is generally reflected in the upper percentiles of the particle size distribution, such as D90 and D99. his is because hard agglomerates are relatively large and therefore primarily affect the right tail of the size distribution rather than the distribution as a whole. Consequently, lower percentile metrics such as D10 or D50 may not adequately capture the formation of hard agglomerates. Therefore, our model focuses on predicting the upper-percentile particle sizes, specifically D90 and D99 in this study. As shown in Fig.~\ref{fig:Exp59_Validation}C, our model can accurately predict the evolution of D90 and D99. The particle size initially increases as agglomeration dominates. Subsequently, the particle size gradually decreases due to deagglomeration. At the end of the process, both D90 and D99 approach their initial values, indicating that there is no hard agglomerate in the system. As explained in Section \ref{sec:Model-PBM-Kernels}, hard agglomerates irreversibly alter the particle size distribution, while soft agglomerates tend to deagglomerate back into their original forms. Therefore, a system containing only soft agglomerates does not exhibit any net change in the final particle size. Hence, the results from both the model predictions and experimental data for Case 1 are reasonable and consistent with the complete insolubility of KCl in hexane, which prevents the formation of hard agglomerates.

\begin{figure}[ht!]
\centering
    \includegraphics[scale=1]{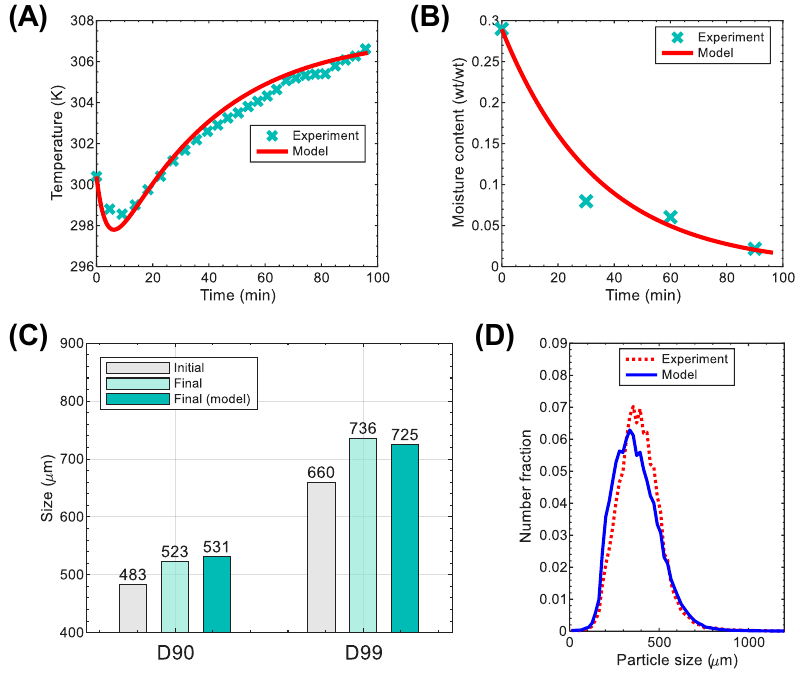}
    \caption {Comparison between the experimental data and model prediction for the system consisting of ASA and IPA (Case 2). Panel (A) shows the product temperature. Panel (B) shows the moisture content. Panel (C) compares the values of D90 and D99 at the initial and final times. Panel (D) shows the PSD at the final time.} 
    \label{fig:Exp78_Validation}      
\end{figure}

For Case 2 (ASA + IPA), the simulated temperature and moisture content exhibit similar trends to those in Case 1 and closely algin with the experimental data (Figs.~\ref{fig:Exp78_Validation}AB). There is a slight drop in the temperature at the beginning due to evaporative cooling. More importantly, the model can correctly predict D90 and D99 (Fig.~\ref{fig:Exp78_Validation}C). Unlike in Case 1, where D90 and D99 return to their initial values at the end of the process, the final values of D90 and D99 in Case 2 are significantly higher than their initial values, which indicates the formation of hard agglomerates. This result is physically logical as ASA is highly soluble in IPA, promoting the formation of hard agglomerates. The model can also accurately estimates the final PSD (Fig.~\ref{fig:Exp78_Validation}D).

To summarize, our model can describe the formation of soft and hard agglomerates properly. The model accurately predicts the evolution of the temperature, moisture content, and particle size, in particular D90 and D99. The model predictions are physically reasonable across different mixtures and operating conditions. 

\subsection{Model-based analysis, design, and optimization}
With the model validated, this section demonstrates the applications of our model in providing further insights into the agitated filter drying processes as well as process design and optimization.

\begin{figure}[ht!]
\centering
    \includegraphics[scale=.75]{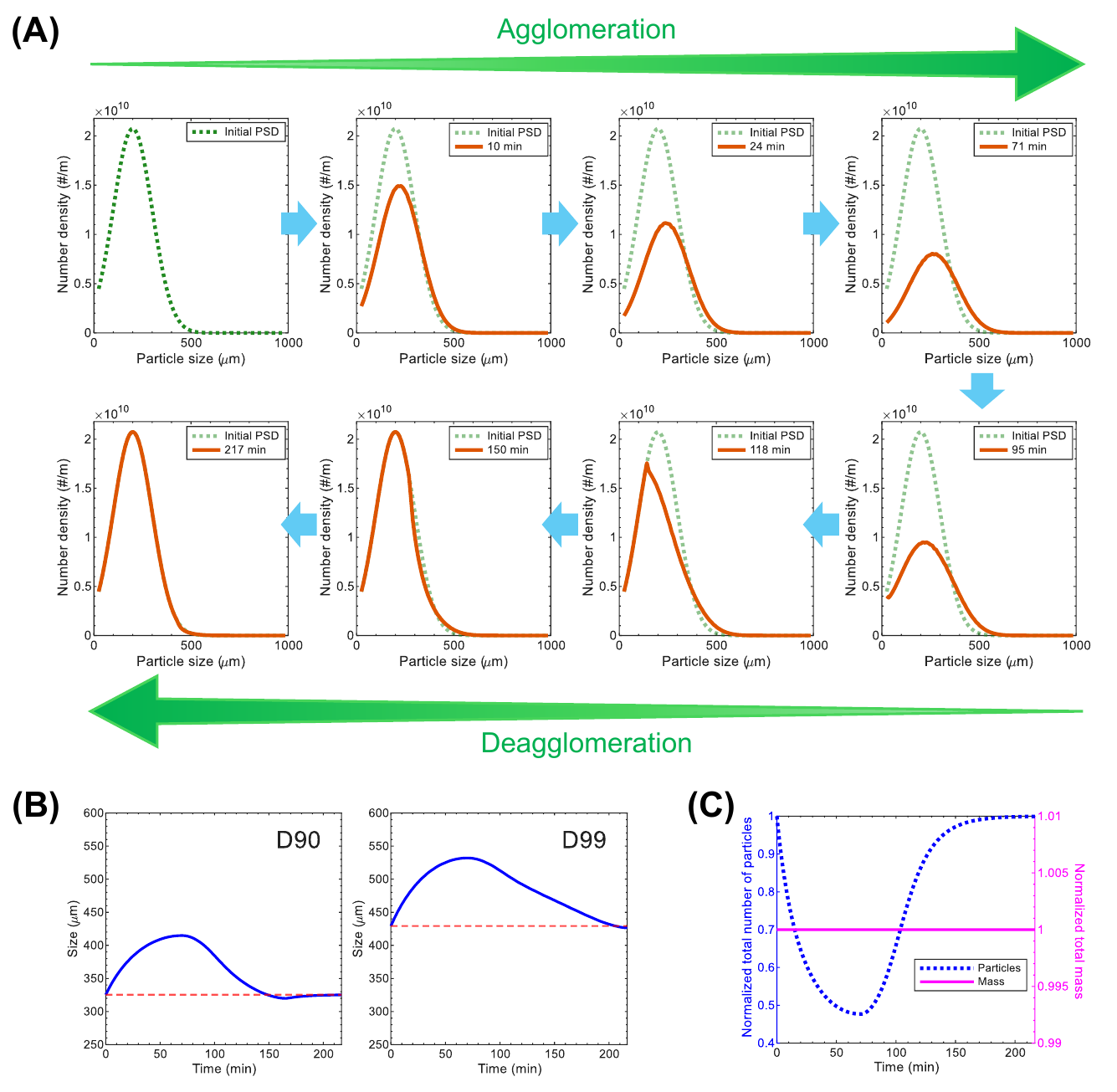}
    \caption {Time evolution of the (A) total mass and number of particles, (B) $D_{90}$ and $D_{99}$, and (C) PSD during AFD, without the formation of hard agglomerates.} 
    \label{fig:SimResult_SoftAgg}      
\end{figure}

First, we compare two different scenarios of AFDs. The first scenario assumes that the solubility of the solute is 0, such that no hard agglomerates can form because no dissolved solute is available for solid bridge formation, similar to Case 1 in Section \ref{sec:ModelValidation}. As shown in Fig.~\ref{fig:SimResult_SoftAgg}A, the PSD initially shifts to the right as agglomeration dominates. At later times, as deagglomeration becomes dominant, the PSD shifts back toward the left. Notably, the final PSD at 217 min is identical to the initial PSD, indicating that no hard agglomerates remain in the system. This behavior is also reflected in the upper percentiles of the PSD, where both D90 and D99 initially increase and subsequently decrease back to their initial values (Fig.~\ref{fig:SimResult_SoftAgg}B). Similarly, the total number of particles initially decreases due to agglomeration and eventually returns to its initial value as the soft agglomerates deagglomerate, while the total particle mass remains conserved throughout the process (Fig.~\ref{fig:SimResult_SoftAgg}C).

\begin{figure}[ht!]
\centering
    \includegraphics[scale=.75]{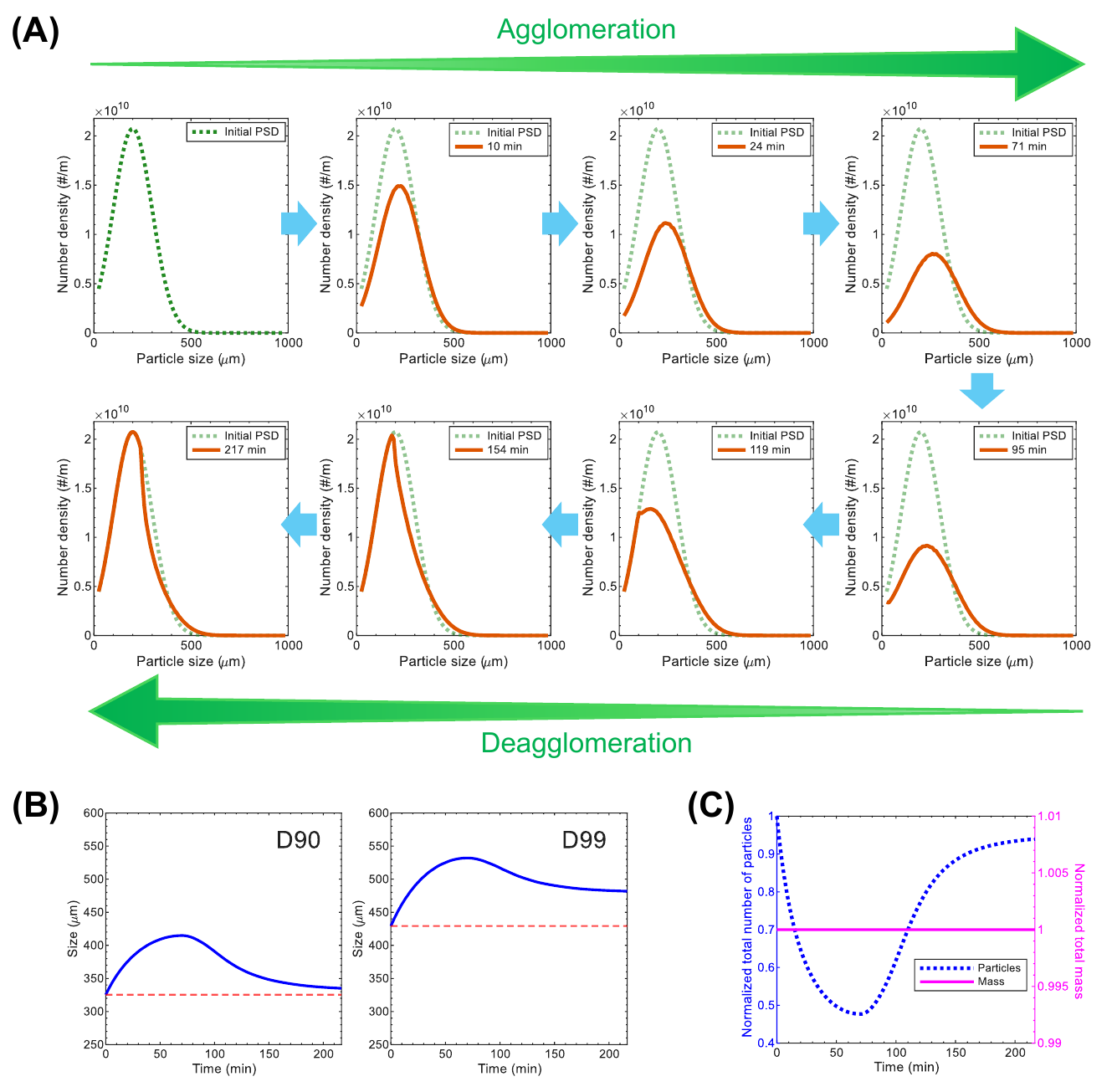}
    \caption {Time evolution of the (A) total mass and number of particles, (B) $D_{90}$ and $D_{99}$, and (C) PSD during AFD, with the formation of hard agglomerates.} 
    \label{fig:SimResult_HardAgg}      
\end{figure}

The second scenario assumes that the solute is highly soluble in the solvent, thereby promoting the formation of hard agglomerates, similar to Case 2 in Section \ref{sec:ModelValidation}. The PSD first shifts toward larger particle sizes when agglomeration dominates and then moves back toward smaller sizes as deagglomeration becomes more significant. In contrast to the first process, however, the final PSD is not identical to its initial value, as evidenced by a slight mismatch between the orange and green curves at 217 min. This behavior is also apparent in the upper percentiles of the PSD, with the final D90 and D99 remaining noticeably higher than their original values (Fig.~\ref{fig:SimResult_HardAgg}B). In addition, the total number of particles does not return to its initial value at the end of the process because the formation of hard agglomerates leads to an irreversible decrease in the total particle number (Fig.~\ref{fig:SimResult_HardAgg}C). Nevertheless, the total particle mass remains conserved throughout the process.

\begin{figure}[ht!]
\centering
    \includegraphics[scale=.75]{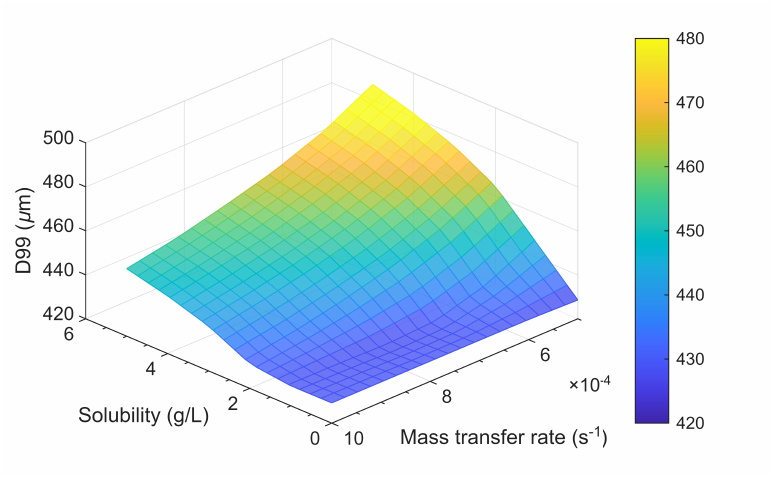}
    \caption {Three-dimensional surface showing the values of D99 at the end of the agitated filter drying process.} 
    \label{fig:SimResult_Param}      
\end{figure}

Finally, we demonstrate how the model can be used to guide the design and optimization of AFDs. The ability to predict the formation of soft and hard agglomerates, as well as the evolution of particle size under different conditions, is valuable for identifying optimal operating conditions and material properties that lead to the desired final product quality. For example, Fig.~\ref{fig:SimResult_Param} shows the values of D99 predicted by our model at different solubilities and mass transfer rates. An increase in the solubility helps promote the formation of hard agglomerates, as can be observed from larger D99 values. On the other hand, an increase in the mass transfer rate can accelerate the disappearance of liquid bridges and thus reduce the probability of successful solid bridge formation. Consequently, D99 decreases higher mass transfer rates. The model can quantitatively capture these competing phenomena and provide further insight into their effects on the particle size. Such information is crucial for improving the design and operation of AFDs to mitigate undesired agglomeration, thereby addressing one of the key challenges in agitated filter drying.

\section{Conclusion} \label{sec:Conclusion}
This article presents a novel mechanistic model for describing the formation of soft and hard agglomerates during agitated filter drying. The model integrates heat transfer, mass transfer, and population balance modeling to capture particle agglomeration and breakage throughout the drying process. The model is validated against two sets of experimental data representing distinct agglomeration behaviors: a limiting case involving KCl and hexane, in which only soft agglomerate exist, and a practical case involving ASA and IPA, in which hard agglomerates (undesired) are significant. In both cases, the model accurately predicts the evolution of product temperature, moisture content, and particle size, particularly the upper percentiles D90 and D99 that are most relevant to the formation of hard agglomerates. The validated model is demonstrated for system analysis and parametric studies, enabling model-based design and optimization to minimize the formation of hard agglomerates. Overall, the proposed model provides a mechanistic and quantitative framework for improving the understanding, design, and operation of AFDs, addressing the key challenge associated with undesired agglomeration.

Future work could further extend the applications of this framework beyond those considered in this study. For example, integration of the model with state estimation and model predictive control could enable real-time process monitoring and control of the particle size evolution, providing additional opportunities to improve process performance and product quality.

\section*{Data and Code Availability}  \label{ch2-sec:code}
All software and data will be made available upon publication of the manuscript.

\section*{Acknowledgments} 
This research was supported by Takeda Development Center Americas, Inc.


\section*{Appendices}
\appendix
\renewcommand\thefigure{\thesection.\arabic{figure}} 
\renewcommand\theequation{\thesection.\arabic{equation}}

\setcounter{figure}{0} 
\setcounter{equation}{0} 

\section{Model and Code Verification} \label{app:A}
Population balance modeling has been studied and employed extensively in the literature, but its numerical implementation remains nontrivial, especially when agglomeration and breakage are modeled simultaneously, and thus requires careful consideration. Hence, this section extensively performs model and code verification to ensure that the model equations and numerical scheme are implemented correctly.

\subsection{Comparison with analytical solutions} \label{sec:analytical}
\begin{figure}[ht!]
\centering
    \includegraphics[scale=.9]{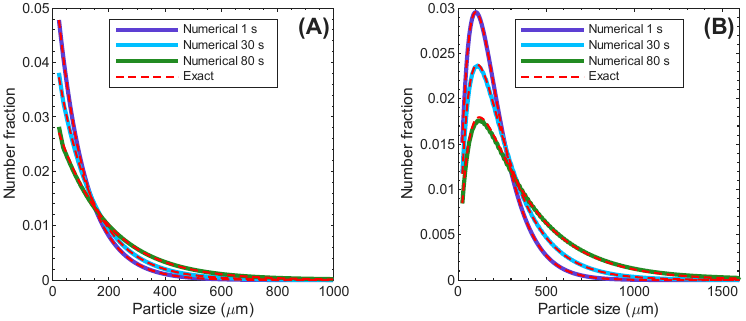}
    \caption {Comparison between the numerical solutions of our model and the analytical solutions for (A) Case A and (B) Case B.} 
    \label{fig:Analytical_Solution}      
\end{figure}

First, our numerical solutions are compared with the analytical solutions obtained from \cite{Scott1968Analytical} for simplified case studies. The population balance equation considered in this comparison includes an agglomeration term with a constant kernel. Two different initial PSDs are selected: an exponential-like PSD (denoted as Case A) and a Gaussian-like PSD (denoted as Case B). The internal coordinate $z$ represents the particle mass (or volume). As shown in Fig.~\ref{fig:Analytical_Solution}, the numerical solutions are nearly identical to the analytical solutions in both the size and time domains. 

\subsection{Error analysis} \label{sec:error}
\begin{figure}[ht!]
\centering
    \includegraphics[scale=.9]{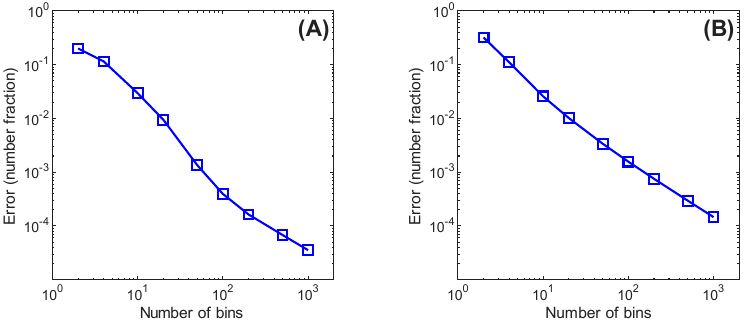}
    \caption {Errors between the numerical solutions of our model and the analytical solutions for (A) Case A and (B) Case B measured at $t = 100$ s.} 
    \label{fig:Analytical_Error}      
\end{figure}

From Cases A and B in Section \ref{sec:analytical}, we analyze the error (root mean square error, RMSE) between the numerical and analytical solutions at different bin sizes. The errors decrease with an increase in the number of bins, which is expected from a proper discretization scheme (Fig.~\ref{fig:Analytical_Error}).  

\subsection{Physical constraint satisfaction} \label{sec:massbalance}
\begin{figure}[ht!]
\centering
    \includegraphics[scale=.9]{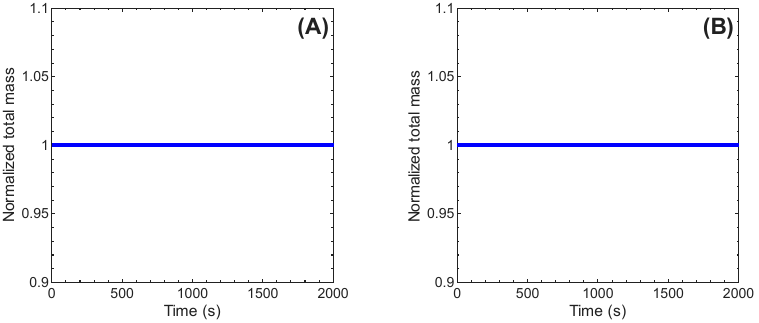}
    \caption {Time evolution of the total mass of all particles for (A) Case A and (B) Case B.} 
    \label{fig:Analytical_Mass}      
\end{figure}

During agglomeration and breakage, the total mass of all particles in the system must be conserved at all times. A correctly implemented population balance model must always generate results that demonstrate mass conservation up to machine precision; even a small deviation indicates that there are errors in the implementation. Based on Cases A and B in Section \ref{sec:analytical}, our numerical solutions satisfy the mass conservation constraint exactly (Fig.~\ref{fig:Analytical_Mass}). 

Results from the studies in this appendix show that our numerical solutions agree with the analytical solutions and satisfy the physical constraint exactly. Reasonable error analysis results are also obtained.

\bibliographystyle{elsarticle-num}
\bibliography{reference}

@book{Smith2018Thermo,
    title={Introduction to Chemical Engineering Thermodynamics},
    author={J. M. Smith and H. C. {Van Ness} and M. M. Abbott and M. T. Swihart},
    editions = {Eight},
    year={2018},
    address = {New York},
    publisher={McGraw-Hill Education}
}

@book{Seader2007Separation,
    title={Separation Process Principles—Chemical and Biochemical Operations},
  author    = {Seader, J. D. and Henley, E. J. and Roper, D. K.},
  edition   = {3rd},
  year      = {2011},
  publisher = {John Wiley \& Sons},
  address   = {Hoboken}
}

@inproceedings{Eldeen1979FallingRateDrying,
  author       = {Sharaf-Eldeen, Yahya I. and Hamdy, M. Y. and Blaisdell, J. L.},
  title        = {Falling rate drying of fully exposed biological materials: A review of mathematical models},
  booktitle    = {Winter Meeting of the American Society of Agricultural Engineers},
  year         = {1979},
  pages = {6522-6543}
}

@article{Ertekin2017ThinLayerReview,
  author  = {Ertekin, Can and Firat, M. Ziya},
  title   = {A comprehensive review of thin-layer drying models used in agricultural products},
  journal = {Critical Reviews in Food Science and Nutrition},
  year    = {2017},
  volume  = {57},
  pages   = {701-717},
  doi     = {10.1080/10408398.2014.910493}
}

@article{Siles2018DryingModel,
  author    = {Siles, J. A. and Mart{\'i}n, M. A. and Molina, E. and Mart{\'i}n, A.},
  title     = {Kinetics of drying inorganic spheres: Simultaneous modeling of moisture and temperature during the constant and falling rate periods},
  journal   = {Drying Technology},
  year      = {2018},
  volume    = {36},
  pages     = {1186-1199},
  doi       = {10.1080/07373937.2017.1391840}
}

@book{Mills1995HeatTransfer,
    title={Heat and Mass Transfer},
    author={Mills, A.},
    dois = {10.4324/9780203752173},
    editions = {First},
    year={1995},
    address = {New York},
    publisher={Routledge}
}

@article{Gnanenthiran2025AFDReview,
  title   = {Undesired agglomeration in agitated filter dryers: A critical review},
  author  = {Gnanenthiran, Suruthi and Hewitt, Christopher and Rao, Pari and Pitt, Kate and Litster, James D. and Smith, Rachel M.},
  journal = {Powder Technology},
  volume  = {466},
  pages   = {121425},
  year    = {2025},
  doi     = {10.1016/j.powtec.2025.121425},
  issn    = {0032-5910}
}

@article{Nichols2002AgglomerateAggregate,
  title   = {A Review of the Terms Agglomerate and Aggregate with a Recommendation for Nomenclature Used in Powder and Particle Characterization},
  author  = {Nichols, Gary and Byard, Stephen and Bloxham, Mark J. and Botterill, Joanne and Dawson, Neil J. and Dennis, Andrew and Diart, Valerie and North, Nigel C. and Sherwood, John D.},
  journal = {Journal of Pharmaceutical Sciences},
  volume  = {91},
  number  = {10},
  pages   = {2103--2109},
  year    = {2002},
  doi     = {10.1002/jps.10191}
}

@article{Sahni2013AFDPerformance,
  title   = {Systematic Investigation of Parameters Affecting the Performance of an Agitated Filter-Dryer},
  author  = {Sahni, Ekneet Kaur and Bogner, Robin H. and Chaudhuri, Bodhisattwa},
  journal = {Journal of Pharmaceutical Sciences},
  volume  = {102},
  number  = {7},
  pages   = {2198--2213},
  year    = {2013},
  doi     = {10.1002/jps.23572}
}

@article{Kougoulos2011AgitatedDrying,
  title   = {Impact of agitated drying on the powder properties of an active pharmaceutical ingredient},
  author  = {Kougoulos, E. and Chadwick, C. E. and Ticehurst, M. D.},
  journal = {Powder Technology},
  volume  = {210},
  number  = {3},
  pages   = {308--314},
  year    = {2011},
  doi     = {10.1016/j.powtec.2011.03.041}
}

@article{Tamrakar2016Dynamic,
  title   = {Dynamic agglomeration profiling during the drying phase in an agitated filter dyer: Parametric investigation and regime map studies},
  author  = {Tamrakar, Ashutosh and Gunadi, Alfeno and Piccione, Patrick M. and Ramachandran, Rohit},
  journal = {Powder Technology},
  volume  = {303},
  pages   = {109--123},
  year    = {2016},
  doi     = {10.1016/j.powtec.2016.09.012}
}

@article{Ottoboni2020Experiment,
title = {Understanding effect of filtration and washing on dried product: Paracetamol case study},
journal = {Powder Technology},
volume = {366},
pages = {305-323},
year = {2020},
issn = {0032-5910},
doi = {10.1016/j.powtec.2020.02.064},
author = {S. Ottoboni and M. Simurda and S. Wilson and A. Irvine and F. Ramsay and C.J. Price},
}

@article{Pena2017SphericalAgglomeration,
  author    = {Ramon Peña and Christopher L. Burcham and Daniel J. Jarmer and Doraiswami Ramkrishna and Zoltan K. Nagy},
  title     = {Modeling and optimization of spherical agglomeration in suspension through a coupled population balance model},
  journal   = {Chemical Engineering Science},
  year      = {2017},
  volume    = {167},
  pages     = {66-77},
  doi       = {10.1016/j.ces.2017.03.055}
}

@article{Patruno2009BreakageKernels,
  author  = {Patruno, L. E. and Dorao, C. A. and Svendsen, H. F. and Jakobsen, H. A.},
  title   = {Analysis of breakage kernels for population balance modelling},
  journal = {Chemical Engineering Science},
  year    = {2009},
  volume  = {64},
  number  = {3},
  pages   = {501--508},
  doi     = {10.1016/j.ces.2008.09.029}
}

@article{Bertin2016Numerical,
title = {Population balance discretization for growth, attrition, aggregation, breakage and nucleation},
journal = {Computers \& Chemical Engineering},
volume = {84},
pages = {132-150},
year = {2016},
doi = {10.1016/j.compchemeng.2015.08.011},
author = {Diego Bertin and Ivana Cotabarren and Juliana Piña and Verónica Bucalá},
}

@article{Scott1968Analytical,
  title   = {Analytic Studies of Cloud Droplet Coalescence I},
  author  = {Scott, William T.},
  journal = {Journal of the Atmospheric Sciences},
  volume  = {25},
  pages   = {54--65},
  year    = {1968},
  doi     = {10.1175/1520-0469(1968)025<0054:ASOCDC>2.0.CO;2}
}

@article{zhang2023extracting,
  title={Extracting particle size distribution from laser speckle with a physics-enhanced autocorrelation-based estimator (PEACE)},
  author={Zhang, Qihang and Gamekkanda, Janaka C and Pandit, Ajinkya and Tang, Wenlong and Papageorgiou, Charles and Mitchell, Chris and Yang, Yihui and Schwaerzler, Michael and Oyetunde, Tolutola and Braatz, Richard D and others},
  journal={Nature communications},
  volume={14},
  number={1},
  pages={1159},
  year={2023},
  publisher={Nature Publishing Group UK London}
}

@article{Lim2016FilterAgglomeration,
  title   = {Understanding and Preventing Agglomeration in a Filter Drying Process},
  author  = {Lim, Hong Lee and Hapgood, Karen P. and Haig, Brian},
  journal = {Powder Technology},
  volume  = {300},
  pages   = {146--156},
  year    = {2016},
  doi     = {10.1016/j.powtec.2016.03.003}
}

@article{Nere2012DryingOptimization,
  author  = {Nere, Nandkishor K. and Allen, Kimberley C. and
             Marek, James C. and Bordawekar, Shailendra V.},
  title   = {Drying Process Optimization for an {API} Solvate Using
             Heat Transfer Model of an Agitated Filter Dryer},
  journal = {Journal of Pharmaceutical Sciences},
  year    = {2012},
  volume  = {101},
  number  = {10},
  pages   = {3886--3895},
  doi     = {10.1002/jps.23237}
}

@article{Sahni2012DryingPerformance,
  author  = {Sahni, Ekneet and Hallisey, Jim and Morgan, Brian and
             Strong, John and Chaudhuri, Bodhisattwa},
  title   = {Quantifying Drying Performance of a Filter Dryer:
             Experiments and Simulations},
  journal = {Advanced Powder Technology},
  year    = {2012},
  volume  = {23},
  number  = {2},
  pages   = {239--249},
  doi     = {10.1016/j.apt.2011.03.002}
}

@article{Sahni2013ContactDryingModel,
  author  = {Sahni, Ekneet Kaur and Chaudhuri, Bodhisattwa},
  title   = {Numerical Simulations of Contact Drying in Agitated
             Filter-Dryer},
  journal = {Chemical Engineering Science},
  year    = {2013},
  volume  = {97},
  pages   = {34--49},
  doi     = {10.1016/j.ces.2013.04.025}
}

@article{Tamrakar2020DEM,
  author  = {Tamrakar, Ashutosh and Zheng, Alex and
             Piccione, Patrick M. and Ramachandran, Rohit},
  title   = {Investigating Particle-Level Dynamics to Understand
             Bulk Behavior in a Lab-Scale Agitated Filter Dryer
             ({AFD}) Using Discrete Element Method ({DEM})},
  journal = {Advanced Powder Technology},
  year    = {2020},
  volume  = {31},
  number  = {1},
  pages   = {477--492},
  doi     = {10.1016/j.apt.2019.11.004}
}

@article{Belekar2022AFDModel,
  author  = {Belekar, Viraj V. and Murphy, Eric J. and
             Heindel, Theodore J. and Nere, Nandkishor K. and
             Subramaniam, Shankar},
  title   = {Modeling Multiphase Heat and Mass Transfer in an
             Agitated Filter Dryer by Integrating Experiment,
             Computations, and Analytical Solutions},
  journal = {Pharmaceutical Research},
  year    = {2022},
  volume  = {39},
  number  = {9},
  pages   = {1971--1990},
  doi     = {10.1007/s11095-022-03382-z}
}

@article{Irndorfer2026DigitalTwin,
  author  = {Irndorfer, S. and Remmelgas, J. and Jajcevic, D. and
             Derrick, A. and Shier, A. and Jolin, G. and
             Glasser, B. and Khinast, J. and Mustakis, J.},
  title   = {A Digital Twin of a Vacuum Filter-Bed Dryer},
  journal = {International Journal of Pharmaceutics},
  year    = {2026},
  volume  = {687},
  pages   = {126427},
  doi     = {10.1016/j.ijpharm.2025.126427}
}

@article{Togni2025Attrition,
  author  = {Togni, Riccardo and Saurer, Eric M. and Hicks, William and
             Rao, Pari and Alabanza, Lady Mae and Clements, Peter and
             DiPietro, Andrew and Derdour, Lotfi and Engstrom, Joshua and
             Hartmanshenn, Clara and Jayaraman, Saivenkataraman and
             Jones-Salkey, Owen and Lamberto, David J. and Mustakis, Jason and
             Ncube, Gqwetha and Kloss, Christoph},
  title   = {A Two-Dimensional Population Balance Model for Predicting
             the Attrition of Elongated Particles During Agitated Drying},
  journal = {Powder Technology},
  year    = {2025},
  volume  = {464},
  pages   = {121198},
  doi     = {10.1016/j.powtec.2025.121198}
}

@article{Ouchiyama1974Granulation,
  author  = {Ouchiyama, Norio and Tanaka, Tatsuo},
  title   = {Mathematical Model in the Kinetics of Granulation},
  journal = {Industrial \& Engineering Chemistry Process Design and Development},
  year    = {1974},
  volume  = {13},
  number  = {4},
  pages   = {383--389},
  doi     = {10.1021/i260052a015}
}

@article{Liu2018PredictivePBM,
  author  = {Liu, Huolong and O'Connor, Thomas and Lee, Sau and Yoon, Seongkyu},
  title   = {A Process Optimization Strategy of a Pulsed-Spray Fluidized Bed
             Granulation Process Based on Predictive Three-Stage Population
             Balance Model},
  journal = {Powder Technology},
  year    = {2018},
  volume  = {327},
  pages   = {188--200},
  doi     = {10.1016/j.powtec.2017.12.070}
}

@article{Hayashi2019PBM,
  author  = {Hayashi, Kentaro and Watano, Satoru},
  title   = {Novel Population Balance Model for Granule Aggregation and
             Breakage in Fluidized Bed Granulation and Drying},
  journal = {Powder Technology},
  year    = {2019},
  volume  = {342},
  pages   = {664--675},
  doi     = {10.1016/j.powtec.2018.10.036}
}

@article{Barrasso2015PBMDEM,
  author  = {Barrasso, Dana and Ramachandran, Rohit},
  title   = {Multi-Scale Modeling of Granulation Processes: Bi-Directional
             Coupling of {PBM} with {DEM} via Collision Frequencies},
  journal = {Chemical Engineering Research and Design},
  year    = {2015},
  volume  = {93},
  pages   = {304--317},
  doi     = {10.1016/j.cherd.2014.04.016}
}

@article{Hayashi2020PBMDEMCFD,
  author  = {Hayashi, Kentaro and Nakamura, Hideya and Watano, Satoru},
  title   = {Numerical Study on Granule Aggregation and Breakage in Fluidized
             Bed Granulation by a Novel {PBM} with {DEM--CFD} Coupling Approach},
  journal = {Powder Technology},
  year    = {2020},
  volume  = {360},
  pages   = {1321--1336},
  doi     = {10.1016/j.powtec.2019.11.027}
}

\end{document}